\documentclass[prb,twocolumn,floatfix,nofootinbib,superscriptaddress]{revtex4-2}
\usepackage{epsfig}
\usepackage{amsmath,amssymb}
\usepackage{bm}
\usepackage{braket}
\usepackage{graphicx}
\usepackage[dvipsnames,usenames]{xcolor}
\usepackage[normalem]{ulem}
\usepackage{soul}
\usepackage{dcolumn}
\usepackage{multirow}
\usepackage{hyperref}
\usepackage{natbib}
\usepackage{siunitx}
\usepackage{algorithm2e}
\usepackage{mathtools}

\hypersetup{
    colorlinks,
    bookmarksopen,
    bookmarksnumbered,
    citecolor=red,
    linkcolor=red,
    urlcolor=red,
}
\emergencystretch=\maxdimen
\RestyleAlgo{ruled}
\SetKwComment{Comment}{/* }{ */}

\newcolumntype{d}{D{.}{.}{-1}} 

\newcommand{\op}[1]{\hat{#1}}
\newcommand{\tvar}{\sigma_{\mathrm{t}}^{2}}

\begin{document}

\title{Emergence of steady quantum transport in a superconducting processor} 

\newcommand{\zju}{School of Physics, ZJU-Hangzhou Global Scientific and Technological Innovation Center, \\and Zhejiang Key Laboratory of Micro-nano Quantum Chips and Quantum Control, Zhejiang University, Hangzhou, China}
\newcommand{\smt}{Science, Mathematics and Technology Cluster, Singapore University of Technology and Design, 8 Somapah Road, 487372 Singapore}
\newcommand{\epd}{Engineering Product Development Pillar, Singapore University of Technology and Design, 8 Somapah Road, 487372 Singapore}
\newcommand{\cqt}{Centre for Quantum Technologies, National University of Singapore 117543, Singapore}
\newcommand{\majulab}{MajuLab, CNRS-UNS-NUS-NTU International Joint Research Unit, UMI 3654, Singapore}
\newcommand{\henan}{Henan Key Laboratory of Quantum Information and Cryptography, Zhengzhou 450001, China}
\newcommand{\sichuan}{College of Physics and Electronic Engineering, and Center for Computational Sciences, Sichuan Normal University, Chengdu 610068, China}
\newcommand{\ioa}{Institute of Automation, Chinese Academy of Sciences, Beijing 100190, China}

\author{Pengfei Zhang} 
\thanks{These authors contributed equally.}
\affiliation{\zju}

\author{Yu Gao} 
\thanks{These authors contributed equally.} 
\affiliation{\zju}

\author{Xiansong Xu} 
\thanks{These authors contributed equally.}
\affiliation{\smt}
\affiliation{\sichuan} 

\author{Ning Wang} 
\affiliation{\zju}

\author{Hang Dong}
\affiliation{\zju}

\author{Chu Guo} 
\affiliation{\henan} 

\author{Jinfeng Deng}
\affiliation{\zju}
\author{Xu Zhang}
\affiliation{\zju}
\author{Jiachen Chen}
\affiliation{\zju}
\author{Shibo Xu}
\affiliation{\zju}
\author{Ke Wang}
\affiliation{\zju}
\author{Yaozu Wu}
\affiliation{\zju}
\author{Chuanyu Zhang}
\affiliation{\zju}
\author{Feitong Jin}
\affiliation{\zju}
\author{Xuhao Zhu}
\affiliation{\zju}
\author{Aosai Zhang} 
\affiliation{\zju}
\author{Yiren Zou}
\affiliation{\zju}
\author{Ziqi Tan}
\affiliation{\zju}
\author{Zhengyi Cui}
\affiliation{\zju}
\author{Zitian Zhu}
\affiliation{\zju}
\author{Fanhao Shen}
\affiliation{\zju}
\author{Tingting Li}
\affiliation{\zju}
\author{Jiarun Zhong}
\affiliation{\zju}
\author{Zehang Bao}
\affiliation{\zju}

\author{Liangtian Zhao}
\affiliation{\ioa}

\author{Jie Hao}
\email{jie.hao@ia.ac.cn}
\affiliation{\ioa}

\author{Hekang Li}
\affiliation{\zju}
\author{Zhen Wang}
\affiliation{\zju}
\author{Chao Song}
\affiliation{\zju}
\author{Qiujiang Guo}
\affiliation{\zju}

\author{H. Wang}
\email{hhwang@zju.edu.cn}
\affiliation{\zju}

\author{Dario Poletti}
\email{dario\_poletti@sutd.edu.sg}
\affiliation{\smt}
\affiliation{\epd}
\affiliation{\cqt}
\affiliation{\majulab}

\begin{abstract}  
Non-equilibrium quantum transport is crucial to technological advances ranging from nanoelectronics to thermal management. 
In essence, it deals with the coherent transfer of energy and (quasi-)particles through quantum channels between thermodynamic baths. 
A complete understanding of quantum transport thus requires the ability to simulate and probe macroscopic and microscopic physics on equal footing. 
Using a superconducting quantum processor, we demonstrate the emergence of non-equilibrium steady quantum transport by emulating the baths with qubit ladders and realising steady particle currents between the baths.
We experimentally show that the currents are independent of the microscopic details of bath initialisation, and their temporal fluctuations decrease rapidly with the size of the baths, emulating those predicted by thermodynamic baths.
The above characteristics are experimental evidence of pure-state statistical mechanics and prethermalisation in non-equilibrium many-body quantum systems. 
Furthermore, by utilising precise controls and measurements with single-site resolution, we demonstrate the capability to tune steady currents by manipulating the macroscopic properties of the baths, including filling and spectral properties. 
Our investigation paves the way for a new generation of experimental exploration of non-equilibrium quantum transport in strongly correlated quantum matter.
\end{abstract}

\maketitle

\section{Main}
Recent advances in quantum thermodynamics and nano-scale devices have shown the increasing importance of quantum transport. 
However, despite its widespread interest and technological significance, a complete knowledge of the microscopic mechanisms that lead to the emergence of steady quantum transport remains elusive. 
The main obstacle resides in the presence of diverse scales: the microscopic motion of the individual particles being transported, and the statistical behaviour of particles in the thermodynamic baths. 
Although new insights into the quantum origin of thermodynamic equilibrium have been given by the eigenstate thermalisation hypothesis~\cite{Deutsch1991, Srednicki1994, RigolOlshanii2008} and typicality~\cite{PopescuWinter2006, GoldsteinZanghi2006}, these approaches are challenged in the investigation of steady quantum transport due to the presence of transport channels. 
This motivates the development of new theoretical, numerical, and experimental frameworks that can capture multiscale non-equilibrium physics. 

\begin{figure*}[!htbp]
    \includegraphics[width=\textwidth]{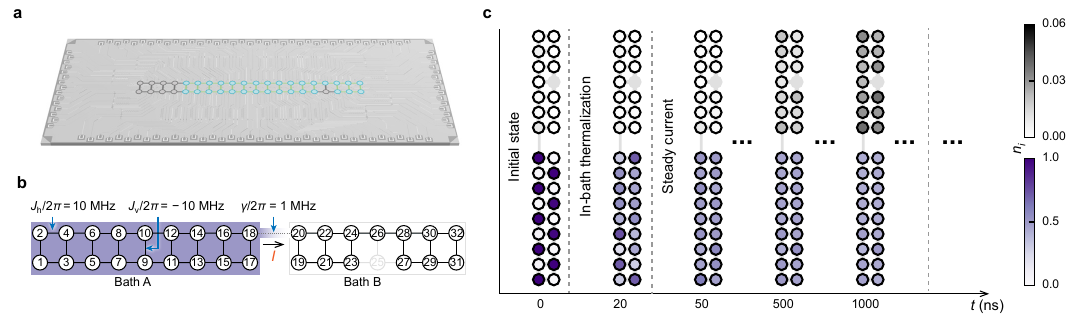}
    \caption{\textbf{Schematics of setup and measurement of site-resolved dynamics. }
        \textbf{a}, Schematic of the device where qubits are depicted as circles with connecting lines representing nearest-neighbour tunable couplings. 31 functional qubits used in the experiment are coloured in cyan.
        \textbf{b}, Mapping to the particle-transport model which comprises two non-integrable baths connected by a weak link between sites $18$ and $20$.
        \textbf{c}, Measured dynamics of the site-resolved particle occupations. Shown are the simultaneously measured excited-state probabilities of individual qubits in both baths (see colorbars on the far right) at a few selected time spots during the dynamics.
        \label{fig:setup}
    }
\end{figure*}

A promising direction is to use quantum simulators such as ultracold atoms~\cite{BrantutEsslinger2012, HusmannBrantut2015, BluvsteinLukin2022}, trapped ions~\cite{SchindlerBlatt2013}, and superconducting qubits~\cite{ GiazottoPekola2006, GiazottoMartinez-Perez2012, MiAbanin2024, RonzaniPekola2018, XuDeng2023}. 
Quantum simulators have the potential to surpass classical computers in investigating many-body quantum dynamics, as they can naturally handle the exponential complexity inherent in these systems~\cite{BernienLukin2017, ChenPan2021, GongPan2021, YaoGuo2023}. 
However, the study of non-equilibrium transport requires additional efforts on these experimental platforms due to the presence of baths. 
Firstly, the description of the bath state often relies on fictitious statistical ensembles, instead of experimentally realistic pure states. 
These ensembles are represented with complex mixed states that are impossible to generate strictly through unitary evolution of a pure state~\cite{SagastizabalDiCarlo2021}.
Secondly, engineering a configurable and scalable bath to have desired macroscopic properties and spectral profiles is still challenging with either analogue or digital quantum simulators~\cite{PekolaKarimi2023, ZhuMonroe2020, SagastizabalDiCarlo2021}. 
Resolving these issues would significantly enhance our capabilities in simulating and understanding complex non-equilibrium quantum dynamics.

In this article, we demonstrate the emergence of steady quantum transport from a single pure initial state in a superconducting processor.
By preparing a population difference between the baths, we observe a generation of inter-bath particle current, the main signature of non-equilibrium transport. 
Specifically, we show that the currents obtained are {\it typical}, as they do not depend on the microscopic dynamics of the baths, but only on their conserved quantities. 
Moreover, these currents are {\it steady}, manifested by a significant decrease in their temporal fluctuations as the system size increases.
We emphasise that 31 qubits are sufficient to capture the above observations, implying a fast transition to the regime where statistical mechanics apply. 
This relies on the recent understanding of quantum chaos~\cite{CasatiChirikov1995, Haake2010}, pure-state quantum statistical mechanics~\cite{PopescuWinter2006, DAlessioRigol2016, BorgonoviZelevinsky2016, Deutsch2018, MoriUeda2018, GoldsteinZanghi2006, Reimann2007, BartschGemmer2009}, and prethermalisation dynamics~\cite{BergesWetterich2004, MoeckelKehrein2008, MallayyaRigol2021, GringSchmiedmayer2012, ReimannDabelow2019} which leads to an extendable intermediate time scale to probe steady current properties~\cite{Fialko2015, XuPoletti2022, XuPoletti2023}. 
Additionally, we show that our setup serves as a flexible platform to study quantum transport by demonstrating a high degree of current tunability.

\section{Experimental setup and site-resolved dynamics} 

Our device consists of a ladder-type array of superconducting transmon qubits, where only neighbouring qubits are directly coupled with adjustable coupling strength (Fig.~\ref{fig:setup}\textbf{a}). This configuration naturally models a hard-core Bose-Hubbard ladder with the effective Hamiltonian given by
\begin{align}
    \frac{\op{H}}{\hbar} = & \sum_{\langle i, j \rangle} J_{i, j} (\op{\sigma}_{i}^{+}\op{\sigma}_{j}^{-} + \op{\sigma}_{i}^{-}\op{\sigma}_{j}^{+}) + \sum_{i} h_{i}\op{\sigma}_{i}^{+}\op{\sigma}_{i}^{-},
    \label{eq:device_hamiltonian}
\end{align}
where $\hbar$ is the reduced Planck constant, $\op{\sigma}_{i}^{+}$ ($\op{\sigma}_{i}^{-}$) is the raising (lowering) operator for qubit $i$, $J_{i, j}$ describes the individually tunable coupling strength 
with $\langle i, j \rangle$ running through all nearest-neighbour sites, and $h_{i}$ characterises the local energy associated with each site $i$.
By default, we set $J_{i, j}$ to $J_{\mathrm{h}}/2\pi = 10$~\unit{\mega\hertz} for horizontally adjacent qubits, $J_{\mathrm{v}}/2\pi = -10$~\unit{\mega\hertz} for vertically adjacent qubits, and $h_i/2\pi=0$, unless otherwise specified.
It is worth noting that the Hamiltonian $\op{H}$ is non-integrable and mixing, which conserves the total particle number $\op{N} = \sum_{i} \op{\sigma}_{i}^{+}\op{\sigma}_{i}^{-}$. 

As illustrated in Fig.~\ref{fig:setup}\textbf{b}, by setting $J_{17,19} \approx 0$ and $J_{18,20}$ to a small value, e.g., $J_{18,20}/2\pi=1.0$~\unit{\mega\hertz}, we effectively divide the qubit ladder into two portions, which we refer to as bath A and bath B.
Accordingly, the Hamiltonian can be reformulated as
\begin{align}
    \op{H}_{\mathrm{tot}}= \op{H}_{\mathrm{A}} + \op{H}_{\mathrm{B}} + \hbar \gamma (\op{\sigma}^+_{18}\op{\sigma}^-_{20} + \op{\sigma}^-_{18}\op{\sigma}^+_{20}),
    \label{eq:hamiltonian}
\end{align}
where $\op{H}_{\mathrm{A}}$ and $\op{H}_{\mathrm{B}}$ are given by Eq.~(\ref{eq:device_hamiltonian}) within their subsystems, and $\gamma \equiv J_{18,20}$.
The new Hamiltonian describes two baths that are weakly coupled with an inter-bath coupling $\gamma$, which lays the foundation for our exploration of steady currents and quantum transport. 
In particular, the coupling $\gamma$ leads to inter-bath particle transport, and the particle current can be defined as
\begin{align}
    \label{eq:current}
    I \left(t\right) = \mathrm{i} \gamma \left(\langle\op{\sigma}^+_{18}\op{\sigma}^-_{20} \rangle- \langle\op{\sigma}^-_{18}\op{\sigma}^+_{20} \rangle\right).
\end{align}

To create the particle current, we initialise two baths using Fock states with different fillings, e.g., half-filled in bath A and empty in bath B, so that the energies of both baths are zero, which prevents the formation of an energy or particle current due to the energy difference.
An arbitrary half-filled state in bath A can be obtained by applying desired single-qubit rotations while all qubits are biased at their respective idle frequencies, with all $J_{i, j}$ tuned to around zero. Then we quickly bring qubits in both baths into resonance by matching their frequencies and meanwhile we modify $J_{i, j}$ to suitable values specified by $J_{\mathrm{h}}$, $J_{\mathrm{v}}$, or $\gamma$. These operations adequately set up the stage at time $t=0$ where two baths with programmable initial states are linked by a weak coupling, which would subsequently undergo quantum quench dynamics following Eq.~\ref{eq:hamiltonian}. We are able to measure, at any instant, observables such as local site occupations and particle currents by simultaneously probing the states of all qubits.

With this scheme, we experimentally prepare an initial state with $L=31$ qubits and monitor the subsequent site-resolved dynamics by taking snapshots of the qubit occupations at a few time spots, as shown in Fig.~\ref{fig:setup}\textbf{c}.
We first observe a fast equilibration process from 0 to 50~\unit{\nano\second}, during which bath A approaches an almost uniform site occupation, which is expected since a sufficiently large bath can gain equilibrium within an intrinsic timescale governed by the intra-bath couplings $J_{\mathrm{h}}$ and $J_{\mathrm{v}}$~\cite{KaufmanGreiner2016, NeillMartinis2016}. This fast equilibration is critical for observing a steady current crossing the weak link, since the initial half-filled Fock states are not in equilibrium.
For longer times after 50~\unit{\nano\second}, it is seen that site-resolved occupation in bath A uniformly decreases while that in bath B uniformly increases, indicating the presence of a particle current resulting from the population difference between the baths. 

\begin{figure}[!htbp]
    \includegraphics[width=\columnwidth]{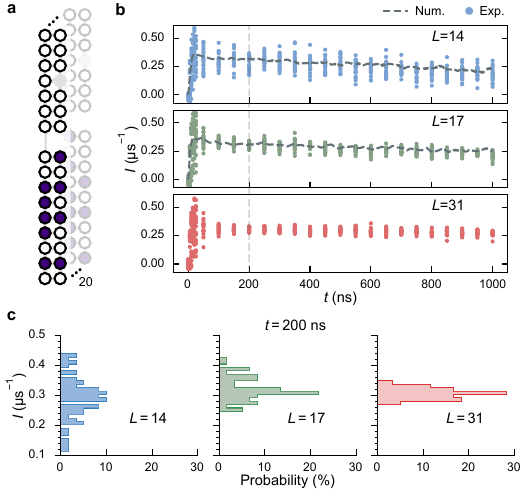}
    \caption{\textbf{Experimental characterisation of the typical currents.}
        \textbf{a}, Schematics of distinct initial states for the investigation of typical currents with a system size of $L=31$. 
        \textbf{b}, Dynamics of the current (circles) for $20$ individual initial states across three distinct system sizes: $L=14$ (blue), $17$ (green), and $31$ (red). 
        The standard error of each current value is noticeably smaller than the marker size, which is not shown for clarity.
        The dashed lines represent the average current values obtained from numerical simulations (unfeasible for the $L=31$ case) across all possible initial Fock states.
        \textbf{c}, Distribution of the current values measured in the experiment at $t=200$~\unit{\nano\second} across $60$ randomly chosen distinct initial Fock states for three system sizes: $L=14$, $17$, and $31$.
    }
    \label{fig:variance}
\end{figure}

\begin{figure*}[!htbp]
    \includegraphics[width=\textwidth]{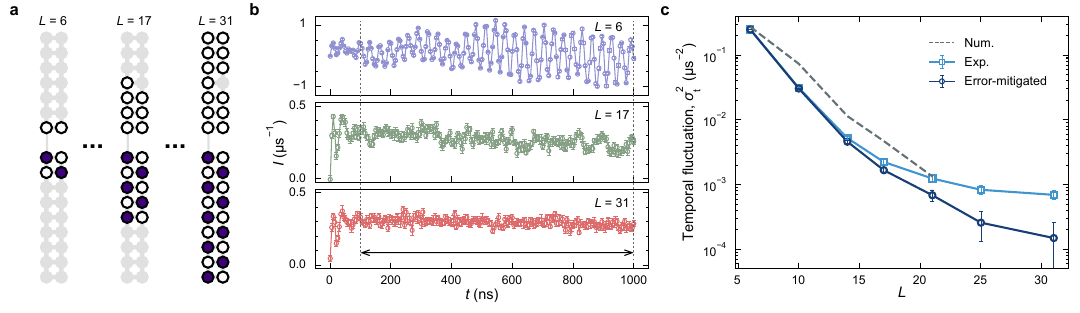}
    \caption{\textbf{Experimental observation of the emergence of steady currents}. \textbf{a}, Schematics of the initial states for three different system sizes: $L = 6$, $17$, and $31$. \textbf{b}, Current dynamics for the depicted initial states in panel \textbf{a} (see Supplementary Section~$4$ for more results on system sizes not shown here). \textbf{c}, Suppressed temporal fluctuations of the current by expanding the system sizes. Fluctuations (light blue squares) are calculated within a time span from $100$~\unit{\nano\second} to $1000$~\unit{\nano\second}, $5$~\unit{\nano\second} per step. The error-mitigated fluctuations are presented with dark blue circles. 
    See Supplementary Section~$4$ for the determination of standard errors (depicted as errorbars).
    Our decoherence-free numerical simulation results (dashed line) serve as a reference, extending up to a system size of $L = 21$. In the Supplementary Section~$5$, we present the numerical results up to $L = 14$ considering the decoherence effect, which yields a closer agreement with the experimental data.
    }
    \label{fig:variance_t}
\end{figure*}

\section{Observation of typical and steady currents}

The behaviours of the current and its dependence on the initial state are closely related to the system size.
To visualise this, we show in Fig.~\ref{fig:variance}\textbf{b} the dynamics of the current for three different system sizes ($L=14$, $17$, and $31$), using $20$ distinct randomly selected initial Fock states (see Fig.~\ref{fig:variance}\textbf{a} for a depiction of $L=31$).
We observe that the experimental results generally follow the numerical ones that are determined from statistical ensembles.
In our numerical simulations, we also consider actual device factors including the inherent cross couplings and the weak anharmonicity of the qubits, causing the numerical simulation for $L=31$ unfeasible (see Supplementary Section~$1$).
We further observe that the current measured at a fixed time, e.g., $I(t=200~\unit{ns})$ demonstrates system-size-dependent fluctuations over different initial states.
For a more detailed examination, we show in Fig.~\ref{fig:variance}\textbf{c} the distribution of the current at $200$~\unit{\nano\second} with $60$ distinct initial states. 
It is seen that the current distribution for $L=31$ is considerably narrower than the other two cases.
This indicates that the vast majority of Fock state initialisation in the half-filling sector will converge towards the same current value at a given time as the system size grows, a manifestation of dynamical typicality~\cite{BartschGemmer2009, NiemeyerGemmer2014, Reimann2018a}. 

The current measured for a single initial state tends to stabilise but also shows system-size-dependent fluctuations within a time span, e.g., from 100 to 1000~\unit{\nano\second}.
Using a single initial state for each system size (Fig.~\ref{fig:variance_t}\textbf{a}), we show the measured current dynamics in Fig.~\ref{fig:variance_t}\textbf{b}.
To evaluate the steadiness of the current, we refer to the temporal fluctuation
\begin{align}
    \tvar &= \frac{1}{K-1}\sum_{k=1}^{K} \left(I\left({t_k}\right) - \frac{1}{K}\sum_{m=1}^{K} I\left({t_m}\right) \right)^{2},
    \label{eq:variance_temporal}
\end{align}
taken within a time span of [100~\unit{\nano\second}, 1000~\unit{\nano\second}] with 5~\unit{\nano\second} per step, and plot its dependence on system size $L$ in Fig.~\ref{fig:variance_t}\textbf{c}. The directly calculated temporal fluctuations first decay exponentially up to $L\approx 17$, and then saturate to a plateau with a small value around $10^{-3}~\unit{\micro\second^{-2}}$, where systematic sampling errors might dominate.
To reduce the effect of sampling errors for revealing the intrinsic scaling behaviour of $\tvar$, we perform an error mitigation analysis based on statistical inference (see Supplementary Section~$4$).
The error-mitigated curve in Fig.~\ref{fig:variance_t}\textbf{c} drops further below the plateau to approximately follow the exponential-decay scaling, indicating the persistence and steadiness of the current for large system sizes (see also Ref.~\cite{XuPoletti2023}).

\section{Tunability of the currents}

\begin{figure*}[!htbp]
    \includegraphics[width=\textwidth]{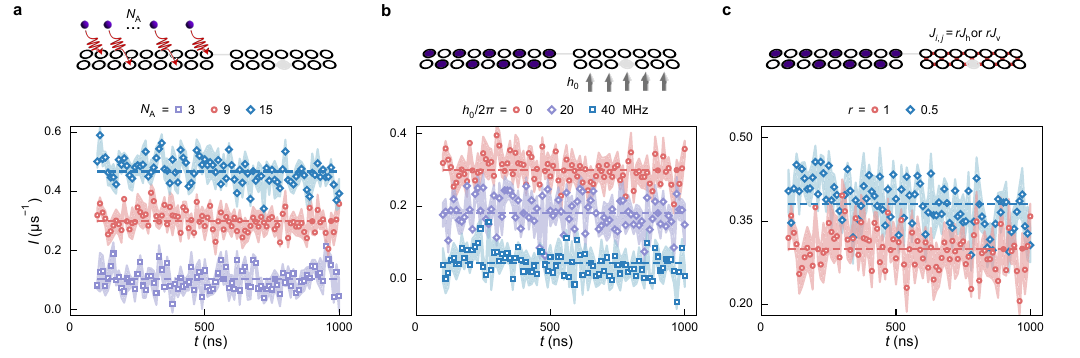}
    \caption{\textbf{Tuning the currents via microscopic control of the baths}. \textbf{a}-\textbf{c}, Diagrams and experimental results for various configurations of the baths, including varying the filling numbers in bath A (\textbf{a}), shifting the on-site particle energies in bath B (\textbf{b}), and adjusting the intra-bath coupling strengths within bath B (\textbf{c}). The shaded regions denote the standard error of the statistical mean. Dashed lines represent the temporal averages of the corresponding data points.
    }
    \label{fig:tunability}
\end{figure*}

The ability to precisely control the baths enables tunability of the currents, an important requirement for the design and understanding of nano-scale devices.  
Here, we show that steady currents can be effectively tuned when the baths are under external controls, indicating the versatility of our protocol.
The most natural tuning mechanism is to adjust the population difference between the baths. 
To show this, we initialise bath A with different fillings from $1/6$ to $5/6$, and measure the resulting currents (Fig.~\ref{fig:tunability}\textbf{a}).
The modulation in population difference leads to a clear separation of the magnitude of the currents.

Another important tuning mechanism relies on the spectral properties of the baths. 
Leveraging on the features of our quantum device, which include selective manipulation of couplings and qubit frequencies, we can systematically modulate the system Hamiltonian and thus directly alter the spectral properties of the baths. 
As the first example, we apply a uniform local potential $h_0$ to all qubits in bath B (see Fig.~\ref{fig:tunability}\textbf{b}) to induce a shift and broadening of the spectrum for bath B, which effectively reduces the overlap between the spectra of the two baths.
Our observations reveal a significant reduction of the particle currents when the local potential $h_0$ increases from $0$~\unit{\mega\hertz} to $40$~\unit{\mega\hertz}. 
Alternatively, we show a different approach in Fig.~\ref{fig:tunability}\textbf{c} by manipulating the intra-bath couplings $J_{i,j}$ within bath B through a scaling factor denoted as $r$ such that $J_{i, j}=r J_{\mathrm{v/h}}$. 
When reducing the couplings by half with $r=0.5$, we observe a $40\%$ increase of the current, which can be explained by the fact that 
the reduction of $r$ implies a spectrum compression of bath B, i.e., the energy required to excite a particle in that bath reduces. 
Such a current increase can also be understood from an equivalent dynamical perspective in which the smaller coupling coefficients slow down the relaxation dynamics of bath B~\cite{L.M.LezamaSantos2023, XuPoletti2022, XuPoletti2023}.

\section{Discussion and Outlook}
 
In this work, we report the very first experimental observation of the emergence of non-equilibrium steady currents between two finite many-body chaotic baths and explore the tunability of the currents.
By integrating insights from quantum information, quantum thermodynamics, and many-body quantum chaos, our study pioneers a new paradigm for experimental investigation of steady quantum transport, and sheds light on the microscopic origin of non-equilibrium statistical physics from bottom-up perspectives. 

Our approach is not only adaptable across various quantum simulation platforms but also offers avenues for further exploration in diverse research directions. 
In future studies, utilising the reconfigurable qubit connectivity in superconducting quantum processors, we can explore more intricate scenarios where two baths are connected via a quantum system emulated by a qubit array. 
In particular, this system can be engineered in situ to have diverse structures and can even be dynamically modulated to advance the investigation of more complex transport phenomena such as rectification~\cite{LandiSchaller2022}, negative differential conductance~\cite{LiLi2012}, and topological currents~\cite{HasanKane2010,AtalaBloch2014}.
Furthermore, the presence of a system between the baths can also facilitate the study of the emergence of nonequilibrium phases and phase transitions~\cite{DallaTorreAltman2010,FitzpatrickHouck2017,BertiniZnidaric2021}. 
Another challenging yet intriguing direction is to characterise the correlation and information propagation within and between the structured baths. 
Specifically, quantifying the information backflow from the baths would help reveal the role of memory effects~\cite{BreuerVacchini2016, deVegaAlonso2017} in quantum transport and other nonequilibrium phenomena.

\vskip 1cm

\noindent {\bf Acknowledgments:} The device was fabricated at the Micro-Nano Fabrication Center of Zhejiang University.  
We acknowledge the support of the National Natural Science Foundation of China (Grants No.~92365301, 12174342, 12274368, 12274367, 12305049, and U20A2076), the National Key Research and Development Program of China (Grant No.~2023YFB4502600), and the Zhejiang Provincial Natural Science Foundation of China (Grant No.~LR24A040002).
D.P. and X.X. are supported by the Ministry of Education Singapore, under the grant MOE-T2EP50120-0019, and by the joint Israel-Singapore NRFISF Research grant NRF2020-NRF-ISF004-3528. 
The authors are grateful to Giuliano Benenti, Gernot Schaller, Hua Yan, Jiaozi Wang, Joel K.W. Yang, Karen Hovhannisyan, Lev Vidmar, Patrycja Łydżba, Ryusuke Hamazaki, Wen-ge Wang, Yuta Sekino, and Zala Lenar\v ci\v c, for fruitful discussions.

\vskip 1cm

\noindent {\bf Author contributions:} X.X. and D.P. proposed the idea; X.X. and C.G. performed the numerical simulation;  P.Z., Y.G., and N.W. carried out the experiments and analyzed the experimental data under the supervision of H.W.; H.L. and J.C. fabricated the device supervised by H.W.; L.Z. and J.H. developed the control electronics; X.X., P.Z., H.W., and D.P. co-wrote the manuscript; All authors contributed to the experimental setup, and/or the discussions of the results and the writing of the manuscript.

\clearpage

\setcounter{table}{0}
\renewcommand{\thetable}{S\arabic{table}}%
\renewcommand{\thefigure}{S$\the\numexpr\value{figure}-4$}%
\setcounter{equation}{0}
\renewcommand{\theequation}{S\arabic{equation}}%
\setcounter{section}{0}
\renewcommand{\thesection}{S\arabic{section}}%
\setcounter{page}{1}
\onecolumngrid
\begin{center}
    \textbf{\large Supplementary Information for ``Emergence of steady quantum transport in a superconducting processor''}\\[.2cm]
\end{center}

\section{Experimental setup} \label{sec:experimental_setup}

Our experiment is performed on a flip-chip superconducting quantum processor containing up to 40 transmon qubits and 58 flux-tunable couplers~\cite{DongPapic2023}.
As depicted in Fig.~1\textbf{a} within the main text, the qubits are arranged in a $20 \times 2$ ladder geometry, and the couplers are used to connect nearest-neighbour qubit pairs both in rungs and side rails.
Each qubit has a control line that integrates both the microwave (XY) control for single-qubit rotations and the flux (Z) control for frequency modulation.
Meanwhile, each coupler has an individual flux (Z) control line to tune the frequency.

\begin{figure*}[ht]
    \includegraphics[width=5in]{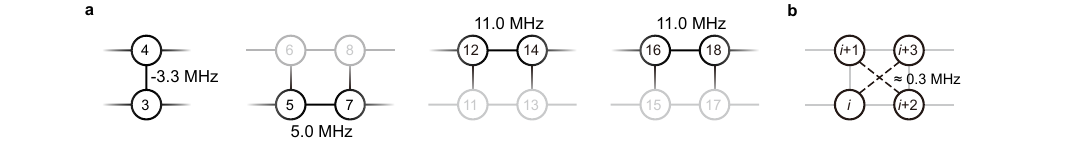}
    \caption{
    \textbf{a}, The nearest-neighbour couplings that deviate from $J_{\mathrm{v}}/2\pi = -10$~\unit{\mega\hertz} or $J_{\mathrm{h}}/2\pi = 10$~\unit{\mega\hertz}. 
    \textbf{b}, Illustration of the cross couplings between next-nearest-neighbour qubits.
    }
    \label{fig:figS_couplings}
\end{figure*}

Two distinct types of couplers are employed to achieve the controllable couplings between horizontally (in side rails) and vertically (in rungs) nearest-neighbour qubits, respectively.
The horizontal couplings can be tuned from approximately $+11.5$~\unit{\mega\hertz} to about $-10.0$~\unit{\mega\hertz} by modulating the frequency of the coupler.
The vertical couplings can be adjusted from approximately $-3.0$~\unit{\mega\hertz} to about $-15.0$~\unit{\mega\hertz}. 
As mentioned in the main text, we set the former to $J_{\mathrm{h}}/2\pi = 10$~\unit{\mega\hertz} and the latter to $J_{\mathrm{v}}/2\pi = -10$~\unit{\mega\hertz}, unless otherwise specified.
For certain couplers, the control of flux and consequently the corresponding coupling between nearest-neighbour qubits is not tunable, as illustrated in Fig.~\ref{fig:figS_couplings}\textbf{a}.
To make two baths weakly coupled via a controllable point contact, we redefine $J_{18, 20}$ as $\gamma$ ($\gamma/2\pi=1$~\unit{\mega\hertz} in the main text), and set $J_{17, 19}/2\pi=0$~\unit{\mega\hertz}.

The spin-$1/2$ $XY$ Hamiltonian in Eq.~(2) in the main text, $\op{H}_{\mathrm{tot}}$, is an idealised representation of the actual experimental setup. 
To give a more precise description, we take into account cross couplings and on-site potentials, yielding the following Hamiltonian
\begin{align}
    \label{eq:device_hamiltonian_bose_hubbard}
    \op{H}_{\mathrm{exp}} &= \op{H}^{\prime}_{\mathrm{tot}} + \op{H}_{\mathrm{x}} + \op{H}_{\mathrm{U}},\\
    \frac{\op{H}_{\mathrm{x}}}{\hbar} &= J_{\mathrm{x}} \sum_{\langle\langle i, j \rangle\rangle } \left(\op{a}_{i}^\dagger \op{a}_{j} + \op{a}_{i} \op{a}_{j}^\dagger \right),\\
    \frac{\op{H}_{\mathrm{U}}}{\hbar} &= \frac{U}{2} \sum_{i} \op{a}_i^\dagger \op{a}_i^\dagger \op{a}_i \op{a}_i,
\end{align}
where $\op{a}_i$ ($\op{a}_i^\dagger$) is the annihilation (creation) operator of the qubit $i$, and $\langle\langle i, j \rangle\rangle$ refers to the next-nearest-neighbour qubit pair $i$ and $j$.
The primary component $\op{H}^{\prime}_{\mathrm{tot}}$ has the same structure as $\op{H}_{\mathrm{tot}}$, except that the spin operators ($\op{\sigma}_i^{-}$ and $\op{\sigma}_i^{+}$) are substituted by the bosonic operators ($\op{a}_i$ and $\op{a}_i^\dagger$). It consists of nearest-neighbour couplings and local potentials of each qubit, as detailed in the main text. 
The second term $\op{H}_{\mathrm{x}}$ accounts for the unintended capacitive cross couplings between next-nearest-neighbour qubit pairs (illustrated in Fig.~\ref{fig:figS_couplings}\textbf{b}).
The value of $J_{\mathrm{x}} / 2\pi \approx 0.3$~\unit{\mega\hertz} ($J_{\mathrm{x}} \approx 0.03J_{\mathrm{h}}$), being significantly smaller than $J_{\mathrm{v/h}}$, exhibits negligible influence on the fast thermalisation dynamics within the baths. 
Moreover, the substantial on-site potential $U/2\pi \approx -175$~\unit{\mega\hertz} ($U \approx -17.5 J_{\mathrm{h}}$) significantly suppresses double occupations at each site. 
However, according to the Schrieffer–Wolff transformation~\cite{BravyiLoss2011, ZhangLai2023}, the large but still finite $U$ results in an effective spin-spin interaction term beyond the spin-$1/2$ $XY$ model.
In the numerical simulation, we use $\op{H}_{\mathrm{exp}}$ rather than $\op{H}_{\mathrm{tot}}$ to take into account the effects of cross couplings and finite on-site potentials. 

\section{Experimental protocol for calibration and measurement} \label{sec:experiment_protocol} 

\begin{figure*}
    \includegraphics[width=\textwidth]{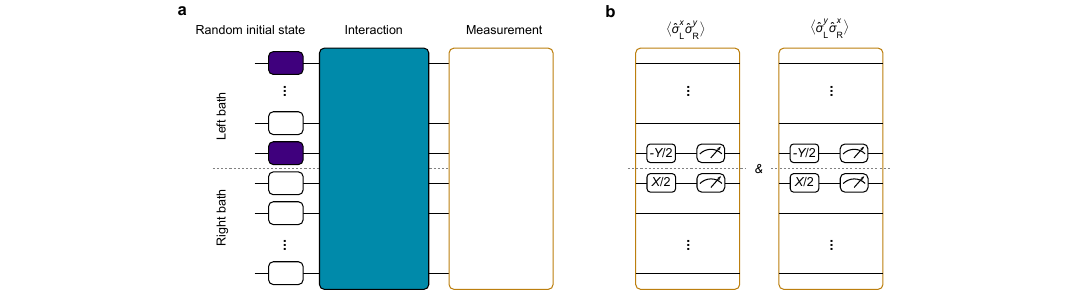}
    \caption{Schematics of the experimental protocol.
    \textbf{a}, The three-step quantum quench including initial state preparation, time evolution, and measurement.
    \textbf{b}, Circuits for the measurement of particle currents.
    }
    \label{fig:figS_current_measurement_protocol}
\end{figure*}

Our investigation of the current dynamics within non-equilibrium quantum many-body systems involves a three-step quantum quench illustrated in Fig.~\ref{fig:figS_current_measurement_protocol}\textbf{a}: 1) initial state preparation, 2) time evolution, and 3) measurement. 
To prepare the initial state, qubits and couplers are set to their idle frequencies, where the interactions between nearest-neighbour qubits are minimised to achieve high-fidelity single-qubit rotations. 
By applying $\pi$ pulses to excite certain qubits into state $\ket{1}$ while preserving the others in state $\ket{0}$, we can generate an arbitrary Fock state within the system.
To evolve the system with the predefined Hamiltonian, we suddenly bias all qubits to the interaction frequency via fast flux pulses. 
Meanwhile, all couplers are tuned to switch on the desired couplings between nearest-neighbour qubits.
After an evolution time $t$, qubits and couplers are tuned back to their respective idle frequencies for measurement of observables (such as particle currents and local occupations).

\subsection{Calibrating the frequencies of qubits and couplers}

This experiment relies on precise control of the frequencies of both qubits and couplers, 
to implement the target local potential ($h_{i}$) and nearest-neighbour couplings ($J_{i, j}$).
The calibration process contains two subroutines: coarse calibration and fine-tuning.
The coarse calibration gives a mapping between the Z pulse amplitude (ZPA) and frequency for each qubit, as well as between the ZPA and corresponding nearest-neighbour coupling strength for each coupler.
The fine-tuning procedures are employed to calibrate and eliminate the deviation of the flux control due to crosstalks and dispersive interactions within the system.

\subsubsection{Coarse calibration of frequencies for qubits and couplers}

\begin{figure*}
    \includegraphics[width=\textwidth]{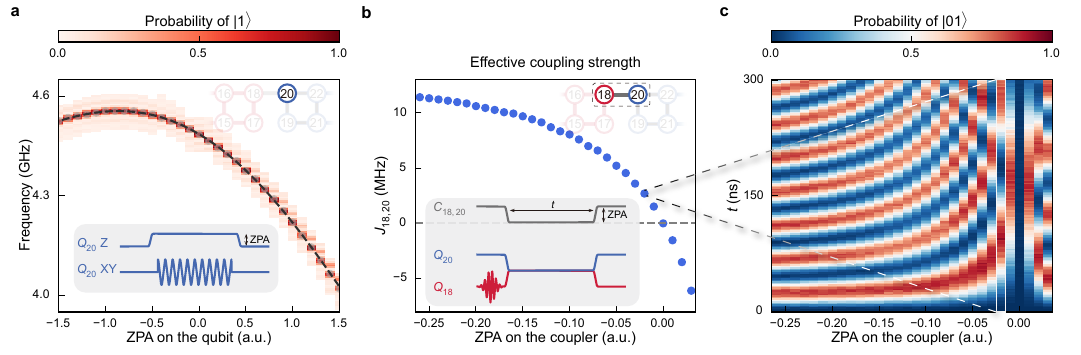}
    \caption{Experimental sequences and results for the coarse calibration of frequencies of qubits and couplers. 
    \textbf{a}, Mapping between the ZPA applied on qubit $20$ (blue circle in the top-right inset) and qubit frequency. The dashed line is a polynomial fitting of this mapping. The bottom inset provides a visual representation of the XY and Z control pulses.
    \textbf{b}, \textbf{c}, Effective coupling strength between qubits $18$ and $20$ as a function of ZPA applied on the corresponding coupler. The mapping in panel \textbf{b} is derived from the observed oscillations in panel \textbf{c}, which illustrates the probability of detecting the two-qubit state in $\ket{01}$. The bottom inset in panel \textbf{b} illustrates the XY and Z control pulses applied to these two qubits along with the Z control pulse applied to the coupler. 
    }
    \label{fig:figS_frequency_calibration}
\end{figure*}

We conduct the qubit spectroscopy experiment to establish a coarse mapping from ZPA to frequency~\cite{YaoGuo2023}. 
The experimental sequence and spectroscopy data are shown in Fig.~\ref{fig:figS_frequency_calibration}\textbf{a}. 
We extract such a mapping with a polynomial fitting for the qubit frequency ranging from $4.0$~GHz to about $4.5$~GHz.

With the sequence displayed in Fig.~\ref{fig:figS_frequency_calibration}\textbf{b}, we calibrate the mapping from the ZPA on the coupler to the effective coupling strength between two nearest-neighbour qubits that are adjacent to the coupler~\cite{ZhangLai2023, DongPapic2023}.
The effective coupling strength (Fig.~\ref{fig:figS_frequency_calibration}\textbf{b}) is determined by fitting the probability oscillation over interaction time (Fig.~\ref{fig:figS_frequency_calibration}\textbf{c}), where the oscillation frequency corresponds to $2|J_{i, j}|$. 
In the case of Fig.~\ref{fig:figS_frequency_calibration}\textbf{b}, a positive ZPA will bring the coupler closer to the qubits, resulting in a negative coupling, while a negative ZPA will result in a positive coupling.

\subsubsection{Fine calibration of qubit frequencies}

\begin{figure*}
    \includegraphics[width=\textwidth]{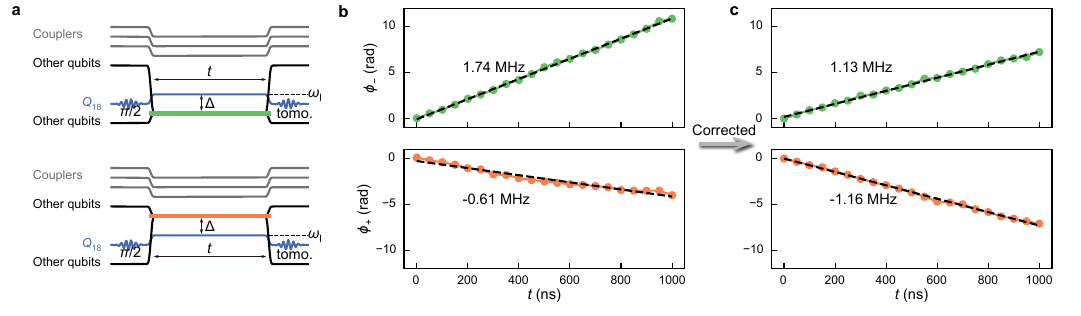}
    \caption{Experimental sequences and results for the fine calibration of qubit frequencies. 
    \textbf{a}, Sequences to determine the qubit frequency shift relative to the interaction frequency $\omega_{\mathrm{I}}$. The target qubit, initialised in a superposition $\left(\ket{0} + \ket{1}\right)/\sqrt{2}$, is biased to $\omega_{\mathrm{I}}$ (blue lines), while the other qubits to $\omega_{\mathrm{I}} \pm \Delta$ (the orange line in the bottom panel and the green line in the top panel, respectively). The phase of the target qubit as a function of time $t$ is measured using tomography pulses. 
    \textbf{b, c}, Experimental results of $\phi_-$ (top panels, green dots) and $\phi_+$ (bottom panels, orange dots). Panels \textbf{b} and \textbf{c} represent the scenarios before and after the convergence of ZPA correction for qubit $20$, respectively. The linear fittings (dashed lines) give the corresponding frequency shifts.
    }
    \label{fig:figS_frequency_calibration_fine}
\end{figure*}

As depicted in Fig.~\ref{fig:figS_frequency_calibration_fine}, we further fine-tune the ZPA of each qubit utilising Ramsey interferometry.
Initially, a $\pi/2$ pulse is applied to prepare the target qubit $i$ in a superposition state.
Using fast Z pulses with ZPAs interpolated from the coarse calibration results, we bias the target qubit to the interaction frequency $\omega_{\mathrm{I}} / 2\pi=4.285$~GHz, the other qubits to $\omega_{-} = \omega_{\mathrm{I}} - \Delta$, and the couplers to their respective operating frequencies. 
In our case, a large detuning $\Delta/2\pi=180$~\unit{\mega\hertz} ($\Delta=18J_{\mathrm{h}}$) is used to effectively isolate the target qubit from the others.
For the target qubit, the phase accumulated relative to the rotating frame of interaction frequency during this process, $\phi_{-}\left(t\right)$, is measured by quantum state tomography.
A similar process, where the other qubits are biased to $\omega_{+} = \omega_{\mathrm{I}} + \Delta$, results in another accumulated phase $\phi_{+}\left(t\right)$.
These two phases ($\phi_{\pm}$) can be expressed as
\begin{align}
    \phi_{\pm}\left(t\right) &= \delta\omega_{\pm}t.
\end{align}
The corresponding frequency shifts ($\delta\omega_{\pm}$) are given by
\begin{align}
    \label{eq:frequency_shift}
    \delta\omega_{\pm} &= \delta\omega^{\mathrm{dis}}_{\pm} + \delta\omega_{0}  ,\\
    \label{eq:dispersive_frequency_shift}
    \delta\omega^{\mathrm{dis}}_{\pm} &= \mp \sum_{j:j\neq i} \frac{J_{i, j}^2}{\Delta},
\end{align}
where $\delta\omega^{\mathrm{dis}}_{\pm}$ characterise the dispersive effects introduced by the other qubits, and $\delta\omega_{0}$ represents the unwanted qubit frequency shift relative to $\omega_{\mathrm{I}}$. 

Taking qubit $20$ as an example,  we can experimentally determine the frequency shifts $\delta\omega_{+} / 2\pi=-0.61$~\unit{\mega\hertz} and $\delta\omega_{-} / 2\pi=1.74$~\unit{\mega\hertz}, as shown in the left two panels of Fig.~\ref{fig:figS_frequency_calibration_fine}\textbf{b}.
This gives $\delta\omega^{\mathrm{dis}}_{\pm}/2\pi=\mp1.17$~\unit{\mega\hertz} and $\delta\omega_{0}/2\pi=0.56$~\unit{\mega\hertz}.
Since the dominant contributions to the dispersive effect come from qubits $18$, $19$, and $22$, we can predict $\delta\omega^{\mathrm{dis}}_{\pm}$ according to Eq.~(\ref{eq:dispersive_frequency_shift}), yielding
\begin{align}
    \delta\omega^{\mathrm{dis}}_{\pm}/2\pi \approx \mp \frac{J_{18, 20}^2 + J_{19, 20}^2 + J_{20, 22}^2}{\Delta} \approx \mp 1.12~\unit{\mega\hertz},
\end{align}
which agree with our experimental findings ($\mp 1.17$~\unit{\mega\hertz}).
The undesired frequency shift $\delta\omega_{0}$ is cancelled by compensating the ZPA on the target qubit. 
Upon correcting the ZPA, two phases ($\phi_{\pm}$) accumulate in the same frequency but with opposite directions, as shown in the right two panels of Fig.~\ref{fig:figS_frequency_calibration_fine}\textbf{b}.
During the interaction process of the quantum quench, the dispersive effect will vanish and hence the fine-tuned ZPA will bring the target qubit to the interaction frequency $\omega_{\mathrm{I}}$.

\subsubsection{Fine calibration of coupler frequencies}

To precisely determine the ZPA of each coupler, we again refer to the oscillation between two-qubit state $\ket{10}$ and $\ket{01}$, similar to the coarse calibration of the coupler frequencies.
However, in this fine-tuning mode, we set the interaction time $t$ as
\begin{align}
    t_{n} &= \left(\frac{1}{2}+n\right) \frac{\pi}{\lvert J_{i, j} \rvert}, \quad n = 0, 1, 2, \cdots.
\end{align}
We use $n=6$ for $\lvert J_{i, j} \rvert/2\pi=10$~\unit{\mega\hertz} and $n=2$ for $\lvert J_{i, j} \rvert/2\pi=1$~\unit{\mega\hertz}.
We then sweep through the ZPA on the target coupler to obtain the optimal ZPA value where the probability of observing two qubits in state $\ket{01}$ is the highest.
During this process, we bias the other qubits to $\omega_{\pm}$ and the couplers to their respective interaction frequencies, similar to the fine calibration of the qubit frequencies. 
The optimal values obtained for $\omega_{\pm}$ are nearly identical, and we use their average as the final ZPA applied to the target coupler.

\vspace{5mm}

Since modifying the coupler frequency will dispersively affect the frequency of two corresponding qubits, the fine-tuning processes for qubits and couplers are performed iteratively.
Usually, the optimal ZPAs for qubits and couplers converge to stable values after two rounds of calibration.
With the above protocol applied to each qubit and coupler, the fine-tuned ZPAs will bring all qubits to $\omega_{\mathrm{I}}$ with desired couplings during the interaction process shown in Fig.~\ref{fig:figS_current_measurement_protocol}\textbf{a}.

\subsection{Measurement of the particle current}

To get the expectation values of the particle current as defined by Eq.~(3) in the main text, an intuitive approach is to reconstruct two-qubit (qubits $18$ and $20$) reduced density matrices through the quantum state tomography.
This method requires the measurement of the quantum state in all nine measurement bases $\{\op{\sigma}^x, \op{\sigma}^y, \op{\sigma}^z\}^{\otimes 2}$~\cite{SteffenMartinis2006, SteffenMartinis2006State, NeeleyMartinis2010, SongPan2017}. 
Alternatively, we can rewrite the particle current in terms of the Pauli operators, 
\begin{align}
    I(t) = -\frac{\gamma}{2} \left(\langle\op{\sigma}_{18}^{x}\op{\sigma}_{20}^{y} \rangle- \langle\op{\sigma}_{18}^{y}\op{\sigma}_{20}^{x} \rangle\right).
     \label{eq:current_pauli_operator}
\end{align}
As a result, the particle current can be determined by measuring the quantum state in two rather than nine measurement bases, giving a simplified measurement protocol as shown in Fig.~\ref{fig:figS_current_measurement_protocol}\textbf{a}, \textbf{b}.

\section{Current dynamics for different inter-bath coupling strengths}
\label{sec:current_different_couplings}

\begin{figure}[!htbp]
    \includegraphics[width=0.5\textwidth]{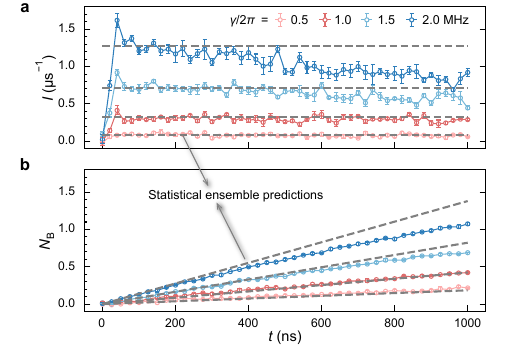}
    \caption{\textbf{Characterisation of dynamics for different inter-bath coupling strengths}.
    \textbf{a},\textbf{b}, The experimental results of the current (\textbf{a}) and the population of bath B (\textbf{b}) with a system size of $L=31$. The corresponding statistical ensemble predictions (grey dashed lines) are obtained from numerical simulations for $L=21$  (see Supplementary Section~$5$).
    For weak inter-bath couplings ($\gamma/2\pi=0.5$, $1.0$~\unit{\mega\hertz}, red lines with circles), the dynamics are in close agreement with the ensemble predictions throughout the process. In contrast, the dynamics develop large deviations from the ensemble predictions during the evolution for strong inter-bath couplings ($\gamma/2\pi=1.5$, $2.0$~\unit{\mega\hertz}, blue lines with circles).
    }
    \label{fig:different_gamma}
\end{figure}

In Fig.~\ref{fig:different_gamma}\textbf{a}, we study the currents under different coupling strengths measured up to $1$~\unit{\micro\second}, with $L=31$ and the same initial state in Fig.~3 of the main text. 
The current driven by the inter-bath couplings will result in a population change in each bath, and the corresponding change in the inter-bath population difference, in turn, alters the magnitude of the current. 
Therefore, it is particularly important to keep the inter-bath coupling sufficiently weak so that the inter-bath bath population differences change mildly throughout the process to produce a steady current.
This exactly falls in the scenario of prethermalisation, where the weak breaking of conserved quantities leads to an intermediate prethermal dynamics (the emergence of a steady current). 
To visualise this, for weak inter-bath couplings ($\gamma/2\pi = 0.5$, $1.0$~\unit{\mega\hertz}), we show in Fig.~\ref{fig:different_gamma} that the current remains almost constant within $1$~\unit{\micro\second}, and the corresponding population of bath B shows a slow and linear increase.
In contrast, for strong inter-bath couplings ($\gamma/2\pi = 1.5$, $2.0$~\unit{\mega\hertz}), there is a clear decay of the current which causes a downward bending in the growth of population. 
To illustrate the differences clearly, the statistical ensemble predictions of thermodynamic baths for both the current and the population, numerically determined by a scaling analysis based on perturbation theory (see Section \ref{subsec:ensemble}), are displayed as references.
In the main text, we consider $\gamma=1.0$~\unit{\mega\hertz} to study the emergence of steady currents since it is sufficiently weak to produce a steady current within $1$~\unit{\micro\second} for $L=31$. 

\section{Analysis of experimental temporal fluctuations} \label{sec:temporal_fluctuation}

In this section, our aim is to perform a detailed analysis of the impact of measurement inaccuracy on both the measured current values and the calculations of corresponding temporal fluctuations as defined by Eq.~(4) in the main text. 
In order to mitigate the impact of finite sampling, we present a sophisticated estimation technique to calculate temporal fluctuations.

\subsection{The statistical framework for the analysis of experimental results}\label{subsec:statistical_framework}

In the following, as mentioned in the main text, the experimental current values are measured at $t$ which is taken from $t_{1} = 100$~\unit{\nano\second} to $t_{K} = 1000$~\unit{\nano\second}, $\Delta t = t_{k+1} - t_{k} = 5$~\unit{\nano\second} per step.
We then model the experimental current value at $t_{k}$ as a sample of the random variable $X_{k}$, whose probability distribution is a normal distribution dependent on two parameters: the population mean $I_{k}$ (or the value averaged from infinite measurement results) and the deviation $\sigma_{k}$ (depends on the measurement sample size $S_{0}$).
In other words,
\begin{align}
    \label{eq:normal_distribution_of_x_for_each_k}
    X_{k} &\sim \mathcal{N}\left(I_{k}, \sigma_{k}^{2}\right), \quad k=1, 2, \cdots, K.
\end{align}
For notational simplicity, we utilise vectors to represent random variables, population means, and standard deviations across different evolution times, i.e.,
\begin{align}
    \boldsymbol{\mathrm{X}} \leftarrow \left(X_{1}, X_{2}, \cdots, X_{K}\right)^{\intercal}, \quad
    \boldsymbol{\mathrm{I}} \leftarrow \left(I_{1}, I_{2}, \cdots, I_{K}\right)^{\intercal}, \quad
    \boldsymbol{\sigma} \leftarrow \left(\sigma_{1}, \sigma_{2}, \cdots, \sigma_{K}\right)^{\intercal}.
\end{align}
We can thus express the probability distribution of $\boldsymbol{\mathrm{X}}$ as a $K$-dimensional multivariate normal distribution,
\begin{align}
    \label{eq:normal_distribution_of_x_as_vector}
    \boldsymbol{\mathrm{X}}
    \sim \mathcal{N}_{K}\left(\boldsymbol{\mathrm{I}}, \boldsymbol{\Sigma}\right),
\end{align}
where the variance-covariance matrix $\boldsymbol{\Sigma} = \boldsymbol{\sigma}^{\intercal} \boldsymbol{\sigma} = \mathrm{diag}\left(\sigma_{1}^{2}, \sigma_{2}^{2}, \cdots, \sigma_{K}^{2}\right)$ is a $K \times K$ diagonal matrix.
The experimental current values can be treated as samples of $\boldsymbol{\mathrm{X}}$, denoted as
\begin{align}
    \left(\boldsymbol{\mathrm{x}}_{1}, \boldsymbol{\mathrm{x}}_{2}, \cdots, \boldsymbol{\mathrm{x}}_{R}\right),
\end{align}
where $R$ is the measurement repetition. 
Additionally, we can define the random variable of the sampling errors as
\begin{align}
    \label{eq:sampling_errors}
    \boldsymbol{\mathrm{E}} = \boldsymbol{\mathrm{X}} - \boldsymbol{\mathrm{I}},
\end{align}
which characterise the differences between the random variable $\boldsymbol{\mathrm{X}}$ and its expected value $\boldsymbol{\mathrm{I}}$.
It will have a multivariate normal distribution with the same variance-covariance matrix as $\boldsymbol{\mathrm{X}}$ but with zero as means, that is,
\begin{align}
    \label{eq:normal_distribution_of_sampling_error_as_vector}
    \boldsymbol{\mathrm{E}} &\sim \mathcal{N}_{K}\left(\mathbf{0}, \boldsymbol{\Sigma}\right),
\end{align}
where $\mathbf{0}$ is the zero vector.
The experimental current values will give us the samples of $\boldsymbol{\mathrm{E}}$ expressed as
\begin{align}
    \boldsymbol{\mathrm{\epsilon}}_{r} = \boldsymbol{\mathrm{x}}_{r} - \boldsymbol{\mathrm{I}}, \quad r=1, 2, \cdots, R.
\end{align}
Another important fact is that the average of $\boldsymbol{\mathrm{x}}_{r}$ and $\boldsymbol{\epsilon}_{r}$ over repetition, represented as
\begin{align}
    \label{eq:tilde_x_and_varepsilon}
    \boldsymbol{\tilde{\mathrm{x}}} = \frac{1}{R} \sum_{r=1}^{R} \boldsymbol{\mathrm{x}}_{r}
    = \boldsymbol{\mathrm{I}} + \boldsymbol{\tilde{\epsilon}},
    \quad
    \boldsymbol{\tilde{\epsilon}} = \frac{1}{R} \sum_{r=1}^{R} \boldsymbol{\epsilon}_{r},
\end{align}
will be samples from
\begin{align}
    \label{eq:normal_distribution_of_averaged_current_and_sampling_error_as_vector}
    \boldsymbol{\tilde{\mathrm{X}}} \sim \mathcal{N}_{K}\left(\boldsymbol{\mathrm{I}}, \frac{1}{R}\boldsymbol{\Sigma}\right),
    \quad
    \boldsymbol{\tilde{\mathrm{E}}} \sim \mathcal{N}_{K}\left(\mathbf{0}, \frac{1}{R}\boldsymbol{\Sigma}\right),
\end{align}
respectively.

\subsection{Standard errors of the experimental current values} \label{subsec:standard_error_of_current}

\begin{figure*}[!htbp]
    \includegraphics[width=\textwidth]{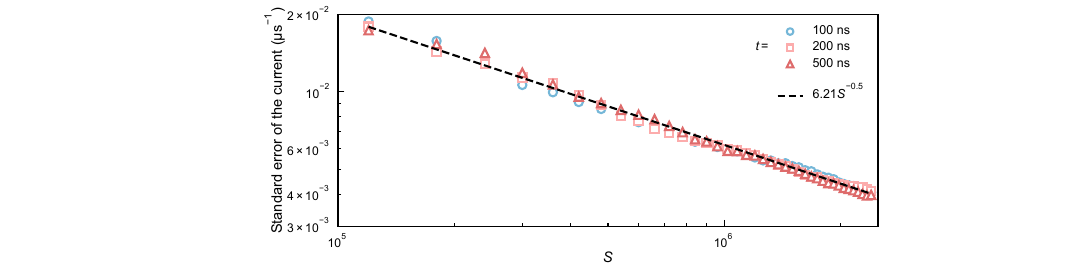}
    \caption{The SEM of the experimental current values as a function of measurement sample size $S$. The dashed line is a fitting of the averaged SEM over three distinct times: $t=100$, $200$, and $500$~\unit{\nano\second}.}
    \label{fig:figS_measurement_standard_error}
\end{figure*}

We start by performing $R=400$ repetitions of current measurements at a given time $t_{1}$, each repetition having a sample size of $S_{0}=6000$.
This measurement process takes about $50$ minutes.
Measurement results are denoted by
\begin{align}
    \left(x_{1, 1}, x_{1, 2}, \cdots, x_{1, 400}\right).
\end{align}
The SEM of $\left(x_{1, 1}, \cdots, x_{1, r}\right)$ will be proportional to $S^{-1/2}$, where $r$ varies from $2$ to $400$ and $S$ is the total sample size $S = rS_{0}$.
In Fig.~\ref{fig:figS_measurement_standard_error}, we plot the SEM as a function of the total sample size $S$ at three distinct times: $t_{1}=100$~\unit{\nano\second}, $t_{21}=200$~\unit{\nano\second}, and $t_{81}=500$~\unit{\nano\second}.
These results suggest that the current values can be measured in the experiment with an SEM less than $4\times10^{-3}~\unit{\micro\second}^{-1}$ when the sample size is greater than $2 \times 10^{6}$.
However, we opt for using smaller sample sizes to achieve a balance between measurement precision and time consumption.
To determine the temporal fluctuation of the current for each system size, we set $S_{0}=6000$ and $R=10$, resulting in a measurement duration ranging from $4$ to $8$ hours depending on the system size. 

\subsection{Calculating the temporal fluctuation from experimental averages} \label{subsec:expression_of_temporal_fluctuation}

\begin{table*}[!htbp]
    \setlength{\tabcolsep}{12pt}
    \renewcommand{\arraystretch}{1.5}
    \centering
    \caption{\label{tab:estimate_of_temporal_fluctuation_and_se} Estimated values of the temporal fluctuation of the current based on different methods and the standard errors.}
    \begin{tabular}{c|*{2}{d}|*{1}{d}}
        \hline \hline
        \multirow{2}{*}{System size, $L$}
        &  \multicolumn{2}{c|}{Temporal fluctuation $\left(\times 10^{-4}~\unit{\micro\second}^{-2}\right)$}
        & \multicolumn{1}{c}{Standard error $\left(\times 10^{-4}~\unit{\micro\second}^{-2}\right)$}
        \\
        & \multicolumn{1}{c}{$\tvar \left(\boldsymbol{\tilde{\mathrm{x}}}\right)$, Eq.~(\ref{eq:temporal_fluctuation_of_x_using_vector})}
        & \multicolumn{1}{c|}{$\widehat{\tvar \left(\boldsymbol{\mathrm{I}}\right)}$, Eq.~(\ref{eq:estimator_of_intrinsic_temporal_fluctuation})}
        & \multicolumn{1}{c}{$\mathrm{SE}\left(\tvar\left(\boldsymbol{\tilde{\mathrm{x}}}\right)\right)$, Eq.~(\ref{eq:se_of_temporal_fluctuation_experimental_values})}
        \\
        \hline
        ~6 & 2491.4 & 2480.8 & 23.3 \\
        10 &  310.5 &  304.6 &  6.4 \\
        14 &   52.3 &   46.0 &  2.7 \\
        17 &   22.4 &   16.7 &  1.8 \\
        21 &   12.4 &    6.9 &  1.4 \\
        25 &    8.3 &    2.6 &  1.3 \\
        31 &    7.0 &    1.5 &  1.1 \\
        \hline
        \hline
    \end{tabular}
\end{table*}

Based on the experimental current values $\left(\boldsymbol{\mathrm{x}}_{1}, \boldsymbol{\mathrm{x}}_{2}, \cdots, \boldsymbol{\mathrm{x}}_{R}\right)$, the most straightforward way to calculate the temporal fluctuation is to use the following expression
\begin{align}
    \label{eq:temporal_fluctuation_of_x_using_vector}
    \tvar \left(\boldsymbol{\tilde{\mathrm{x}}}\right)
    & = \frac{1}{K-1}
    \boldsymbol{\tilde{\mathrm{x}}}^{\intercal}
    \boldsymbol{\mathrm{M}}
    \boldsymbol{\tilde{\mathrm{x}}}
    \nonumber\\
    & = \frac{1}{K-1}\left(
    \boldsymbol{\tilde{\mathrm{x}}}^{\intercal} \boldsymbol{\tilde{\mathrm{x}}}
    - \boldsymbol{\tilde{\mathrm{x}}}^{\intercal} \frac{\boldsymbol{\mathrm{J}}}{K} \boldsymbol{\tilde{\mathrm{x}}}
    \right),
\end{align}
where $\boldsymbol{\mathrm{M}} = \left(\boldsymbol{1} - \boldsymbol{\mathrm{J}}/K \right)$, $\boldsymbol{1}$ is the $K \times K$ identity matrix, and $\boldsymbol{\mathrm{J}}$ is the $K \times K$ matrix of ones. 
This is essentially the same definition as that given in Eq.~(4) of the main text.
The values of $\tvar \left(\boldsymbol{\tilde{\mathrm{x}}}\right)$ for different system sizes are shown in Table~\ref{tab:estimate_of_temporal_fluctuation_and_se} (also depicted as ``Exp.'' in Fig.~3\textbf{c} of the main text).

Taking the expectation of the preceding expression, we have
\begin{align}
    \label{eq:expectation_of_temporal_fluctuation_of_x_using_vector}
    \mathbb{E} \left(\tvar \left(\boldsymbol{\tilde{\mathrm{x}}}\right)\right)
    = \mathbb{E} \left(\frac{1}{K-1}
    \boldsymbol{\tilde{\mathrm{x}}}^{\intercal}
    \boldsymbol{\mathrm{M}}
    \boldsymbol{\tilde{\mathrm{x}}}
    \right)
    &= \mathbb{E}
    \left( \frac{1}{K-1}
    \left(
    \boldsymbol{\mathrm{I}}^{\intercal}\boldsymbol{\mathrm{M}}\boldsymbol{\mathrm{I}}
    + \boldsymbol{\tilde{\epsilon}}^{\intercal}\boldsymbol{\mathrm{M}}\boldsymbol{\tilde{\epsilon}}
    + \boldsymbol{\mathrm{I}}^{\intercal}\boldsymbol{\mathrm{M}}\boldsymbol{\tilde{\epsilon}}
    + \boldsymbol{\tilde{\epsilon}}^{\intercal}\boldsymbol{\mathrm{M}}\boldsymbol{\mathrm{I}}
    \right)
    \right)
    \nonumber\\
    &= \tvar \left(\boldsymbol{\mathrm{I}}\right)
    + \mathbb{E}
    \left( \frac{1}{K-1}
    \left(
    \boldsymbol{\tilde{\epsilon}}^{\intercal}\boldsymbol{\mathrm{M}}\boldsymbol{\tilde{\epsilon}}
    + \boldsymbol{\mathrm{I}}^{\intercal}\boldsymbol{\mathrm{M}}\boldsymbol{\tilde{\epsilon}}
    + \boldsymbol{\tilde{\epsilon}}^{\intercal}\boldsymbol{\mathrm{M}}\boldsymbol{\mathrm{I}}
    \right)
    \right)
    \nonumber\\
    &= \tvar \left(\boldsymbol{\mathrm{I}}\right) + \frac{1}{KR}\boldsymbol{\sigma}^{\intercal}\boldsymbol{\sigma},
\end{align}
which clearly differs from the expected temporal fluctuation $\tvar \left(\boldsymbol{\mathrm{I}}\right)$.
In other words, a direct calculation of the temporal fluctuation using experimental averages is effectively using an estimator with bias and, therefore, will produce a non-negligible error when the expected temporal fluctuation is comparable with the last term in the above expression (known as the bias of the estimator).
This equation also shows that the value of this bias can be further reduced by increasing the number of repetitions $R$, which requires additional time for the measurement of current values.

\subsection{Error mitigation of the temporal fluctuation} \label{subsec:unbiased_estimator_of_fluctuation}

Here, by employing an appropriate estimation protocol, we can circumvent the problem due to finite repetitions ($R$) and obtain significantly improved results.
In particular, we introduce an error-mitigated estimator given by
\begin{align}
    \label{eq:estimator_of_intrinsic_temporal_fluctuation}
    \widehat{\tvar \left(\boldsymbol{\mathrm{I}}\right)}
    = \frac{1}{K-1} \boldsymbol{\tilde{\mathrm{x}}}^{\intercal} \boldsymbol{\mathrm{M}} \boldsymbol{\tilde{\mathrm{x}}}
    - \frac{1}{KR\left(R-1\right)}\sum_{r=1}^{R} \left(\boldsymbol{\mathrm{x}}_{r} - \boldsymbol{\tilde{\mathrm{x}}}\right)^{\intercal} \left(\boldsymbol{\mathrm{x}}_{r} - \boldsymbol{\tilde{\mathrm{x}}}\right).
\end{align}
It can be easily verified that this estimator is unbiased, i.e.,  
\begin{align}
    \mathbb{E}\left(
    \widehat{\tvar \left(\boldsymbol{\mathrm{I}}\right)}
    - \tvar \left(\boldsymbol{\mathrm{I}}\right)
    \right)=0.    
\end{align}
The values given by this estimator are provided in Table~\ref{tab:estimate_of_temporal_fluctuation_and_se} (also depicted as ``Error-mitigated'' in Fig.~3\textbf{c} of the main text).

\subsection{Estimating the variance of sampling errors}\label{sec:variance_of_sampling_errors}

\begin{figure*}[!htbp]
    \includegraphics[width=\textwidth]{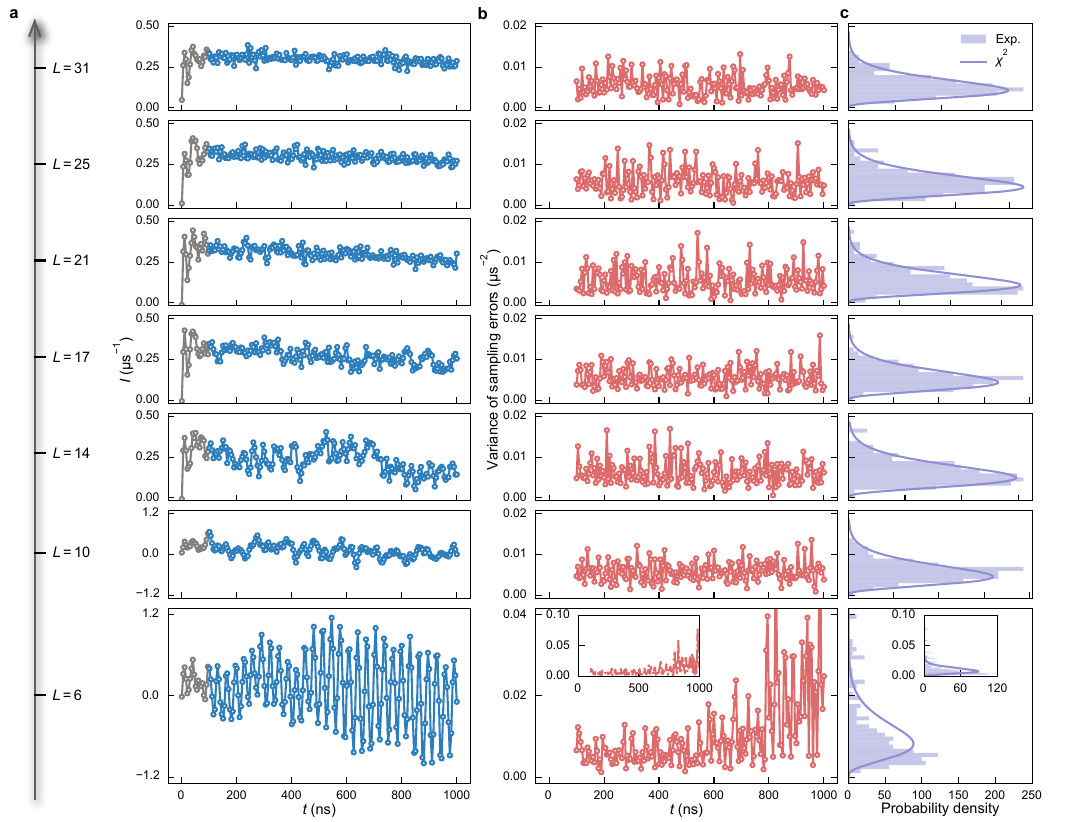}
    \caption{
    {\bf a}, Experimental current values from $0$~\unit{\nano\second} to $100$~\unit{\nano\second} (grey circles) and from $100$~\unit{\nano\second} to $1000$~\unit{\nano\second} (blue circles). {\bf b}, Variances of sampling errors versus evolution time from $100$~\unit{\nano\second} to $1000$~\unit{\nano\second}.
    \textbf{c}, Distributions of variances of sampling errors. Solid purple lines represent the corresponding chi-square distributions. 
    }
    \label{fig:figS_current_measurement_variance_and_distribution}
\end{figure*}

Before we delve into the analysis of standard errors of temporal fluctuation, let us first investigate the property of sampling errors. 
From the definition in the section on statistical models, we note that $\boldsymbol{\mathrm{E}}$ and $\boldsymbol{\mathrm{X}}$ share the same variance-covariance matrix. 
Therefore, an estimator of the variance-covariance matrix of sampling errors can be written as a function of $\boldsymbol{\mathrm{x}}_{r}$, which is
\begin{align}
    \widehat{\boldsymbol{\Sigma}}
    = \frac{1}{R-1} \sum_{r=1}^{R}
    \left(\boldsymbol{\mathrm{x}}_{r} - \boldsymbol{\tilde{\mathrm{x}}}\right)^{\intercal}
    \left(\boldsymbol{\mathrm{x}}_{r} - \boldsymbol{\tilde{\mathrm{x}}}\right).
\end{align}
When we take infinite repetitions, the estimator will give the same value as the original variance-covariance matrix.
Since the variance-covariance matrix $\boldsymbol{\Sigma}$ should be a diagonal matrix, we can focus on the diagonal elements.
Their values are given by
\begin{align}
    \widehat{\Sigma}_{k, k}
    = \frac{1}{R-1}\sum_{r=1}^{R}
    \left(x_{k, r} - \tilde{x}_{k}\right)^{2},
\end{align}
where $\widehat{\Sigma}_{k, k}$ is the $k$-th diagonal element of $\widehat{\boldsymbol{\Sigma}}$, and $x_{k, r}$ ($\tilde{x}_{k}$) the $k$-th element of $\boldsymbol{\mathrm{x}}_{r}$ ($\boldsymbol{\tilde{\mathrm{x}}}_{r}$).
For different system sizes ($L$ ranging from $6$ to $31$), the average current, $\tilde{x}_{k}$, and the corresponding estimate of variance, $\widehat{\Sigma}_{k, k}$, as a function of $t_{k}$ are shown in Fig.~\ref{fig:figS_current_measurement_variance_and_distribution}\textbf{a}, \textbf{b}.

\begin{table*}[!htbp]
    \setlength{\tabcolsep}{8pt}
    \renewcommand{\arraystretch}{1.5}
    \centering
    \caption{\label{tab:mean_sampling_variance} Results of Brown–Forsythe test and mean sampling variances for different system sizes.}
    \begin{tabular}{c|*{3}{d}}
        \hline \hline
        System size, $L$
        & \multicolumn{1}{c}{$p$-value of Brown Forsythe test}
        & \multicolumn{1}{c}{Mean sampling variance, $\bar{\sigma}^{2}$ ($\times 10^{-3}~\unit{\micro\second}^{-2}$)} 
        \\
        \hline
        ~6  & 0.00 & 10.6 \\
        10 &  1.00 &  5.9 \\
        14 &  0.81 &  6.2 \\
        17 &  0.92 &  5.7 \\
        21 &  0.21 &  5.6 \\
        25 &  0.40 &  5.7 \\
        31 &  0.88 &  5.5 \\
        \hline
        \hline
    \end{tabular}
\end{table*}

Our statistical framework can be greatly simplified if all sampling errors have the same variance for a given system size (called homogeneity of variance or homoscedasticity). 
We can verify this homogeneity using the Brown-Forsythe test~\cite{BrownForsythe1974}.
Here, we let $x_{k, r}$ serve as the $r$-th sample of the $k$-th sample group and obtain the $p$-values as shown in Table~\ref{tab:mean_sampling_variance}.
For $L \ge 10$, the $p$-values are statistically significant ($\gg 0.05$), suggesting that these sampling errors may share the same variance.
As depicted in Fig.~\ref{fig:figS_current_measurement_variance_and_distribution}\textbf{c}, for each system size ($L \ge 10$), this property can be visualised by the distribution of estimated variances of sampling errors ($\widehat{\Sigma}_{k, k}$, $k=1, 2, \cdots, K$), which is in close agreement with a corresponding scaled chi-squared distribution ($\chi^{2}$).
The chi-squared distribution has a degree of freedom given by $\left(R-1\right)$ and an average value expressed as
\begin{align}
    \bar{\sigma}^{2} = \frac{1}{K} \sum_{k=1}^{K} \left(\widehat{\Sigma}_{k, k}\right).
    \label{eq:chi_squared_mean}
\end{align}
In the subsequent discussion, we will denote $\bar{\sigma}^{2}$ as the mean sampling variance, whose values for different system sizes ($L$ ranging from $6$ to $31$) are given in Table~\ref{tab:mean_sampling_variance}.
We note that the case of $L=6$ is an exception, as shown in Fig.~\ref{fig:figS_current_measurement_variance_and_distribution}\textbf{b}.
This is likely because the current fluctuates rapidly over $t$ when $t \geq 450$~\unit{\nano\second}, making it significantly more susceptible to noise.

\subsection{Standard error of the temporal fluctuation calculated from experimental averages}

With the above discussion, we can now investigate the standard error of the temporal fluctuation given by Eq.~(\ref{eq:temporal_fluctuation_of_x_using_vector}).
According to the \textit{second-order} Taylor expansion of an arbitrary function $f\left(\boldsymbol{\tilde{\mathrm{X}}}\right)$ around the expected values $\boldsymbol{\mathrm{I}}$, the equation of error propagation is~\cite{PutkoGreen2001, Zhang2006, Skala2021, SipkensGregory2023}
\begin{align}
    \mathrm{SE}_{\text{2nd-order}}\left(f\left(\boldsymbol{\tilde{\mathrm{X}}}\right)\right)
    = \left.
    \sqrt{
    \sum_{k=1}^{K} \left(\frac{\partial f}{\partial \tilde{X}_{k}} \frac{\sigma_{k}}{\sqrt{R}}\right)^{2}
    + \frac{1}{2}\sum_{j=1}^{K}\sum_{k=1}^{K} \left(\frac{\partial^{2} f}{\partial \tilde{X}_{j} \tilde{X}_{k}} \frac{\sigma_{j}}{\sqrt{R}}\frac{\sigma_{k}}{\sqrt{R}}\right)^{2}
    }
    \right|_{\boldsymbol{\tilde{\mathrm{X}}}=\boldsymbol{\mathrm{I}}}.
\end{align}
Given the definition of temporal fluctuation, the partial derivatives are given by
\begin{align}
    \label{eq:partial_derivative_first_order}
    \left.
    \frac{\partial \left(\tvar\left(\boldsymbol{\tilde{\mathrm{X}}}\right)\right)}{\partial \tilde{X}_{k}}
    \right|_{\boldsymbol{\tilde{\mathrm{X}}} = \boldsymbol{\mathrm{I}}}
    & = \frac{2}{K-1}\left(I_{k} - \frac{1}{K}\sum_{j=1}^{K} I_{j}\right),
    \\
    \label{eq:partial_derivative_second_order}
    \left.
    \frac{\partial^{2} \left(\tvar\left(\boldsymbol{\tilde{\mathrm{X}}}\right)\right)}{\partial \tilde{X}_{j} \partial \tilde{X}_{k}}
    \right|_{\boldsymbol{\tilde{\mathrm{X}}} = \boldsymbol{\mathrm{I}}}
    & =
    \begin{dcases*}
       \frac{2}{K},                  & for $j = k$\\
       -\frac{2}{K\left(K-1\right)}, & for $j \neq k$\\
    \end{dcases*},
\end{align}
The uncertainty of $\tvar\left(\boldsymbol{\tilde{\mathrm{X}}}\right)$ will be
\begin{align}
    \label{eq:se_of_temporal_fluctuation_error_propagation}
    \mathrm{SE}_{\text{2nd-order}}
    \left(\tvar\left(\boldsymbol{\tilde{\mathrm{X}}}\right)\right)
    = \frac{\sqrt{2}}{K-1}
    \sqrt{
        \frac{2}{R}\sum_{k=1}^{K} \left(\left(I_{k} - \frac{1}{K}\sum_{j=1}^{K} I_{j}\right)\sigma_{k}\right)^{2}
        + \frac{1}{\left(KR\right)^{2}} \left(\sum_{k=1}^{K} \sigma_{k}^{2}\right)^{2}
        + \frac{K-2}{KR^{2}} \sum_{k=1}^{K} \sigma_{k}^{4}
    }.
\end{align}
The standard error of the temporal fluctuation is influenced not only by the variances of the sampling errors, but also by the statistical characteristics of the expected current values. 

We take the average current values ($\tilde{x}_{1}, \tilde{x}_{2}, \cdots, \tilde{x}_{K}$) and the corresponding estimates of variance ($\widehat{\Sigma}_{1, 1}, \widehat{\Sigma}_{2, 2}, \cdots, \widehat{\Sigma}_{K, K}$) to evaluate the above expression, which gives us
\begin{align}
    \label{eq:se_of_temporal_fluctuation_experimental_values}
    \mathrm{SE}
    \left(\tvar\left(\boldsymbol{\tilde{\mathrm{x}}}\right)\right)
    = \frac{\sqrt{2}}{K-1}
    \sqrt{
        \frac{2}{R} \sum_{k=1}^{K} \widehat{\Sigma}_{k, k} \left(\tilde{x}_{k} - \frac{1}{K}\sum_{j=1}^{K} \tilde{x}_{j}\right)^{2} 
        + \frac{1}{\left(KR\right)^{2}} \left(\sum_{k=1}^{K} \widehat{\Sigma}_{k, k}\right)^{2}
        + \frac{K-2}{KR^{2}} \sum_{k=1}^{K} \left(\widehat{\Sigma}_{k, k}\right)^{2}
    }.
\end{align}
The results given by this expression are provided in Table~\ref{tab:estimate_of_temporal_fluctuation_and_se} (also depicted as errorbars for both the ``Exp.'' and ``Error-mitigated'' temporal fluctuations in Fig.~3\textbf{c} of the main text).

\section{Numerical results of current dynamics}\label{sec:numerical_results}

\begin{figure}
   \centering
    \includegraphics[width=\textwidth]{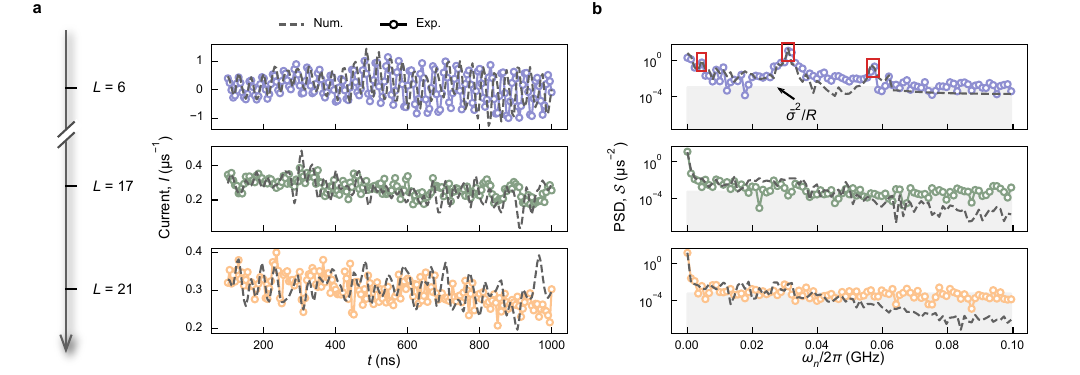}
    \caption{The current dynamics and PSD of the numerical simulations and experimental results. \textbf{a}, The numerical (dashed lines) and experimental (solid lines with circles) current dynamics from Fig.~3 are plotted for different system sizes $L=6$ (purple), $17$ (green), $21$ (orange) from top to bottom, respectively. \textbf{b}, The corresponding power spectrum density of the current dynamics from panel \textbf{a}. The grey-shaded regime represents the noise level (characterised by $\bar{\sigma}^2/R$) of the experimental results. Note that we present the experimental results for $L=21$ to compare with numerical simulations as the $31$-site system is beyond the reach of numerical simulations.}  \label{fig:figS_current_dynamics_and_fft_exp_vs_num}
\end{figure}

In this section, we numerically study the current dynamics by considering the time evolution of the full Hamiltonian $\op{H}_{\mathrm{exp}}$.
Exploiting the $U(1)$ symmetry of the Hamiltonian, i.e., particle number conservation, we work in the symmetry sector with the particle number given by the initial Fock state.
This allows us to deal with a smaller Hilbert space dimension to simulate a larger system size up to $L=21$ (with a local Hilbert space dimension of $3$).
We then employ the fourth-order Runge-Kutta method with a time step of $0.1$~\unit{\nano\second}, while the expectation value of the particle current is calculated with a time step of $1$~\unit{\nano\second}. 

In the following, we first perform a numerical benchmark with the experimental results, followed by a discussion on the impact of dephasing noise. 
Finally, we present the prediction of the current from statistical ensembles.

\subsection{Numerical benchmark with experimental results}\label{subsec:benchmark_bwtween_num_and_exp}
We benchmark the experimental results with numerical simulations for different system sizes ($L=6$, $17$, and $21$).
As depicted in Fig.~\ref{fig:figS_current_dynamics_and_fft_exp_vs_num}\textbf{a}, the numerical and experimental results show consistent dynamics. 
To visualise this more clearly, we analyse the current dynamics in the frequency domain. 
Specifically, we calculate the power spectral density (PSD) by performing the fast Fourier transform on the current dynamics.
The PSD is given by
\begin{align}
    \mathcal{S}\left(\omega_{n}\right) &= \frac{1}{K}\left\lvert\sum_{k=1}^{K}I_{k} \exp{\left(-\mathrm{i} \omega_{n} k \Delta t\right)} \right\rvert^{2},\\
    \omega_{n} &= \frac{2\pi}{K \Delta t}n,
\end{align}
where $n \in \left\{0, 1, \cdots, \left\lceil K/2 \right\rceil \right\}$, $\lceil x \rceil$ is the ceiling function of $x$, and $t_K$ is the time period taken to perform the Fourier transform.
Moreover, we note that the relationship between the PSD and temporal fluctuation is expressed as 
\begin{align}
    \tvar \left(\boldsymbol{\mathrm{I}}\right) = \frac{2}{K}\sum_{n=1}^{\lceil K/2 \rceil} \mathcal{S}\left(\omega_{n}\right).
\end{align}
For $L=6$, both the numerical and experimental PSDs show three distinguishable peaks at frequencies of $\omega_{n}/2\pi=4$, $31$, and $57$~\unit{\mega\hertz}.
The highest peak at $\omega_{n}/2\pi=31$~\unit{\mega\hertz} corresponds to the large oscillations of the current observed in Fig.~\ref{fig:figS_current_dynamics_and_fft_exp_vs_num}\textbf{a}.
In other words, the PSD near $\omega_{n}/2\pi=31$~\unit{\mega\hertz} has a substantial contribution to the temporal fluctuation. 
As the system size grows, the temporal fluctuations decrease as discussed in the main text. 
The corresponding PSD calculated from the numerical current dynamics becomes smoother, and both the low- and high-frequency contributions are suppressed. 
The PSDs derived from experimental results are able to capture the above-mentioned signatures within the low-frequency regime.
The mismatch in the high-frequency regime is due to the sampling errors caused by finite measurement samples (see Section~\ref{sec:variance_of_sampling_errors}).
The sampling errors, in the context of PSD, behave as noise and can be characterised by an average amplitude of $\bar{\sigma}^{2}/R$ (see Eq.~\ref{eq:chi_squared_mean} and Table~\ref{tab:mean_sampling_variance}), as depicted in Fig.~\ref{fig:figS_current_dynamics_and_fft_exp_vs_num}\textbf{b}.
As the system size grows, the PSD amplitudes of the current dynamics in the high-frequency regime gradually decrease to be equal to or even smaller than that of the sampling errors.
Nevertheless, as discussed in Section~\ref{subsec:expression_of_temporal_fluctuation}, by increasing the measurement samples, we can further suppress the sampling errors to eliminate the mismatch between numerical and experimental results. 
More importantly, despite the presence of sampling errors, we can still perform the error mitigation to better estimate the intrinsic value of the temporal fluctuation.

\subsection{Reduction of temporal fluctuation due to dephasing} \label{subsec:dephasing_numerical_temporal_fluctuation}

\begin{figure}
   \centering
    \includegraphics[width=\textwidth]{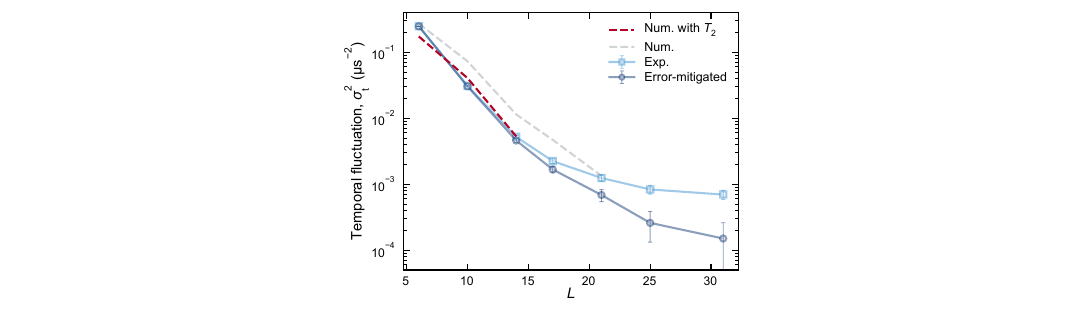}
    \caption{Temporal fluctuation of current dynamics with dephasing noise versus system size (red dashed line). The dephasing noise is parameterised by $T_2 = 7$~\unit{\micro\second}.}
    \label{fig:figS_temporal_fluctuation_withT2}
\end{figure}

In this work, we aimed to study the emergence of the steady current within an isolated quantum system as a fundamental test for the microscopic origin of non-equilibrium statistical mechanics. 
In the previous section, we show a good agreement for current dynamics between the numerical and experimental results. However, in the evaluation of temporal fluctuations for Fig.~3\textbf{c}, 
we notice that the experimental results are systematically smaller than those of the numerical ones for system sizes less than 21. For $L>21$, the experimental results show a converging trend to a plateau when the system size increases.
The latter behaviour can be explained by the variance of sampling errors as discussed in Section~\ref{sec:temporal_fluctuation}. 
The former behaviour can be attributed to noise on the system of interest, introduced by external environments.

To take the noise into account, a common approach is to describe the system dynamics using the Gorini-Kossakowski-Sudarshan-Lindblad master equation,
\begin{align}
    \frac{d \op{\rho}}{dt} = -\frac{\mathrm{i}}{\hbar} \left[\op{H}, \op{\rho}\right] + \sum_{\alpha}  \sum_{j}  \left( \op{L}^\alpha_j \op{\rho} {\op{L}^{\alpha \dagger}_j}  - \frac{1}{2}  \left\{ {\op{L}^{\alpha\dagger}_j}   \op{L}^\alpha_j, \op{\rho} \right\} \right),
\end{align}
where the jump operator $\op{L}_{j}^{\alpha}$ models the noise from the environment.
Here, the superscript $\alpha$ denotes the type of the noise, and the subscript $j$ indicates the specific site that the noise acts upon. 
Two common types of noise are known as amplitude damping noise and dephasing noise, with their jump operators given by 
\begin{align}
    \op{L}^1_j = \sqrt{\frac{1}{T_1}} \op{a}_j, \quad 
    \op{L}^2_j = \sqrt{\frac{2}{T_2}} \op{a}^\dagger_j \op{a}_j = \sqrt{\frac{2}{T_2}}   \op{n}_j,
\end{align} 
respectively. 
The mean value of $T_1$, experimentally estimated as $30$~\unit{\micro\second}, is far larger than the timescale we studied and therefore the amplitude damping is neglected. 
Hence, in the following numerical simulation, we only consider the impact of dephasing with the average value $T_2=7$~\unit{\micro\second}. 
In Fig.~\ref{fig:figS_temporal_fluctuation_withT2}, we present numerical simulation results for various system sizes ranging from $L=6$ to $14$. 
These results suggest that, when compared to the noise-free simulation, dephasing noise reduces temporal fluctuations, leading to a better agreement with the experimental results.

\subsection{Prediction of the current from statistical ensembles} \label{subsec:ensemble}

\begin{figure}
    \centering
    \includegraphics[width=\columnwidth]{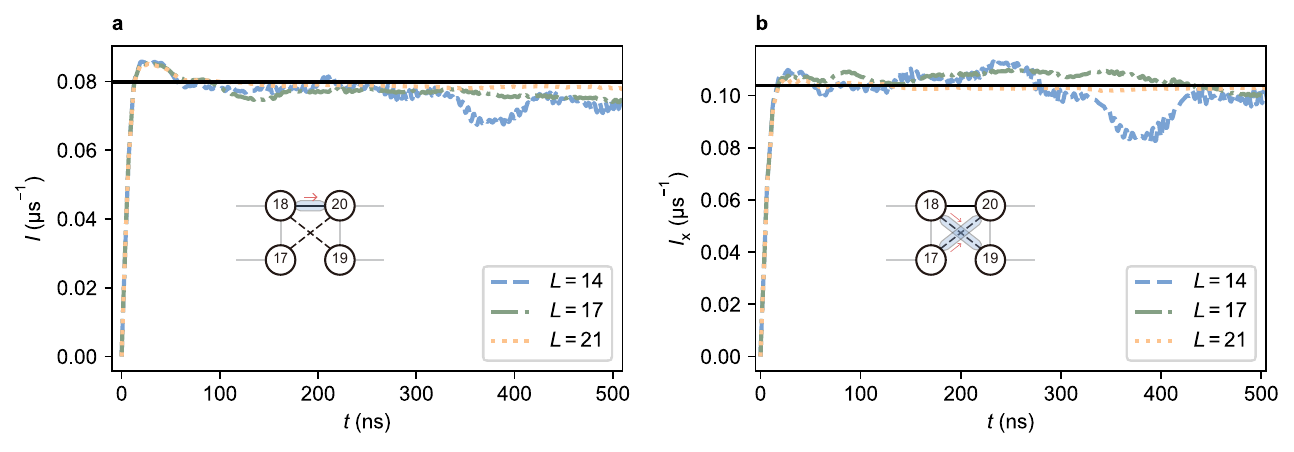}
    \caption{
    Ensemble-based current dynamics for currents through {\bf a}, qubit pair $(18, 20)$ with the inter-bath coupling $\gamma_0 /2 \pi = 0.5$~\unit{\mega\hertz} and {\bf b}, qubit pairs $(17, 20)$ and $(18, 19)$ with the cross coupling $J_{\mathrm{x}} /2 \pi = 0.3$~\unit{\mega\hertz}.  
    The black solid lines indicate the values of $\mathcal{I}_0$ (panel a) and $\mathcal{I_{\mathrm{x}}}$ (panel b), determined by averaging the current values between $60$~\unit{\nano\second} to $150$~\unit{\nano\second} for $L=21$. 
    }
    \label{fig:supp_ensemble_current}
\end{figure}

As discussed in the main text, with an arbitrary typical initial state, our setup can generate steady currents that are consistent with the case where bath A is described by a statistical ensemble.
In our case, since the energy of any possible Fock state $\ket{\bm{s}}$ of bath A is zero, the corresponding ensemble is given by an infinite temperature state.
This state is defined as $\op{\rho}_{\infty}=D^{-1}\sum_{\bm{s}}\ket{\bm{s}}\bra{\bm{s}}$, where $D$ is the number of Fock states in the symmetry sector given by the particle number conservation.
Consequently, in the numerical simulation, we iterate the initial state over all possible Fock states for bath A (while all sites in bath B are fixed to state $\ket{0}$). 
The average current therefore exactly describes the scenario where bath A is prepared in $\op{\rho}_{\infty}$.
In the following, we refer to the average current as the ensemble-based current.

For a weak inter-bath coupling of $\gamma_{0}/2\pi=0.5$~\unit{\mega\hertz}, the ensemble-based current dynamics for different system sizes are shown in Fig.~\ref{fig:supp_ensemble_current}\textbf{a}. 
As the system size grows, the ensemble-based currents quickly converge to a fixed value $\mathcal{I}_{0}$. 
To estimate this value, we take the average of the current values from $t=60$~\unit{\nano\second} to $150$~\unit{\nano\second} for the system size $L=21$. This interval is chosen to avoid the impact of transient dynamics. 
It follows that this average is found to be $\mathcal{I}_{0}=0.08~\unit{\micro\second}^{-1}$,
giving a prediction from the statistical ensemble for a thermodynamically large system. We thus use it as the reference value of steady current for $\gamma/2\pi = 0.5$~\unit{\mega\hertz} in Fig.~\ref{fig:different_gamma}\textbf{a} in the main text. 
For the other values of $\gamma$ used in the main text, we employ the perturbation theory for a scaling analysis over $\gamma$~\cite{XuPoletti2022, XuPoletti2023}, which provides a prediction in the form of $\mathcal{I}_{0} \left(\gamma/\gamma_{0}\right)^{2}$.
However, for a finite system size, we note that this prediction only works within a timescale determined by $\gamma$ in an inverse relationship.
This is because the population difference between two baths changes drastically in the case of a large $\gamma$.
Therefore, in Fig.~\ref{fig:different_gamma}\textbf{a} in the main text, the experimental results for $\gamma/2\pi=1.5$ and $2.0$~\unit{\mega\hertz} deviate from the corresponding predictions, whereas those for $\gamma/2\pi=0.5$ and $1.0$~\unit{\mega\hertz} closely match the corresponding predictions even up to $1$~\unit{\micro\second} or more.

For reference values of the bath B population in Fig.~\ref{fig:different_gamma}\textbf{b} in the main text, it is necessary to take into account the effects of the cross couplings between qubits $17$ and $20$, as well as between qubits $18$ and $19$.
In Fig.~\ref{fig:supp_ensemble_current}\textbf{b}, we show the ensemble-based current due to the cross couplings. 
Similar to the currents introduced by $\gamma_0$, the currents due to the cross couplings quickly converge to $\mathcal{I}_{\mathrm{x}}=0.10~\unit{\micro\second}^{-1}$. 
This value is determined in the same manner as the case of $\mathcal{I}_{0}$.
Consequently, the predictions of bath B population will be
\begin{align}
    N_{\mathrm{right}}\left(t\right) = \left(\mathcal{I}_{0}\left(\frac{\gamma}{\gamma_{0}}\right)^{2} + \mathcal{I}_{\mathrm{x}}\right)t.
\end{align}
For the weak inter-bath coupling cases in Fig.~\ref{fig:different_gamma}\textbf{b} in the main text, the close agreement between the experimental results and this predicted linear growth gives another evidence of steady current behaviour. 
\end{document}